\documentclass[twocolumn,showpacs,preprintnumbers,amsmath,amssymb]{revtex4}
\input epsf
\begin{document}

\title{Energy Dependence of Scattering Ground State Polar Molecules}
\author{Christopher Ticknor$^\dagger$} 
\affiliation{
ARC Centre of Excellence for Quantum-Atom Optics and
Centre for Atom Optics and Ultrafast Spectroscopy,
Swinburne University of Technology, Hawthorn, 
Victoria 3122, Australia}
\date{\today}

\begin{abstract}
We explore the total collisional cross section of ground state polar 
molecules in an 
electric field at various energies, focusing on RbCs and 
RbK.  An external electric field polarizes the molecules and induces  
strong dipolar interactions leading to non-zero partial 
waves contributing to the scattering even as the collision energy goes to 
zero.  This results in the need to compute scattering problems with many 
different values of total M to converge the total cross section.
An accurate and efficient approximate total cross section is introduced and 
used to study the low field temperature dependence.  To understand the 
scattering of the polar molecules we compare a semi-classical cross section
with quantum unitarity limit.  This comparison leads to the ability to 
characterize the scattering based on the value of the electric field and
the collision energy.  General and simple forms of the characteristic electric 
field and energy are given, enabling characterization of the scattering.
\end{abstract}
\maketitle
\section{Introduction}
Recently the ballistic expansion of a Bose-Einstein condensate (BEC) 
of $^{52}$Cr showed the influence of the magnetic dipole-dipole 
interaction\cite{dicold}.
In these experiments an interplay of trap geometry and magnetic polarization 
was used to clearly illustrate the interaction of the dipoles.  
Another example of spin-spin dipolar interactions adding character to 
ultracold matter is the perturbative effect observed in the p-wave 
Feshbach resonances in $^{40}K$ \cite{regal-pwave} and
$^{40}$K-$^{87}$Rb \cite{rbk-pwave} where different $|m_l|$s  
have distinct resonant magnetic field values \cite{CT-pwave}.
These experiments offer a glimpse of the additional character relatively
weak dipole-dipole interactions offer in ultracold atomic systems.
Attention has begun to turn towards polar molecules which have large
electric dipole moments. Theories predict many novel phase 
transitions for dipolar gases \cite{phase1,buchler}.  Further heightening 
the interest in polar molecules are its applications, which range from quantum
computing \cite{qc1,qc2} to tests of fundamental symmetries \cite{fund1,fund2}.

With such remarkable possibilities,
it is not surprising that rapid experimental progress 
should soon produce ultracold ground state polar molecules.
There are many methods of producing cold molecules; for a review  
see Ref. \cite{cold-rev}.  One
of the most exciting techniques used to produce cold polar molecules
is photoassociation (PA).  This method produces cold polar molecules
by binding two distinct ultracold alkali atoms through a series of
optical transitions which ultimately lead to the formation of a
polar molecule such as RbCs \cite{kerman}, KRb \cite{wang} or NaCs \cite{big}.
Recently RbCs was produced in its absolute vibrational ground state at a
temperature of 100 $\mu K$ \cite{sage}.  The production of a cold/dense sample
of these heteronuclear alkali dimers would constitute the
realization of a strongly interacting dipolar system near T=0. 
Considering only the physics of collisions, this system presents
a series of exciting experiments, such as the detection of field
linked states \cite{FL}, the study of ultracold rotationally
inelastic collisions, the study of fully hydrodynamic systems, and 
ultracold chemistry \cite{chem}.

In light of the current experimental progress there is an immediate need 
to understand
the collisions of polar molecules, so that collisional experiments 
can be understood and control of the molecular interactions can be achieved.
Previous theoretical scattering studies looked at how the long range dipolar 
interactions are affected by external fields, first those in
weak-field seeking states \cite{FL,CT-OH,AA-JB-PRA} and later those in
strong-field seeking states \cite{CT-PR}.  Other studies looked at the
ways magnetic and electric fields can be used  to control the molecular 
state in collisions of atom-molecule systems \cite{Krems}.

For ultracold ground state polar molecules in an electric field, only 
long range interactions at extremely cold temperatures have been studied 
previously \cite{CT-PR}. 
That work showed potential resonances (PR) occur with the application of
an external field.  These resonances emerge from the electric field changing 
the long range character of the lowest adiabatic curve, from $1/R^6$ to 
$1/R^3$, adding many bound states to the system.  This mechanism leads to 
broad semi-regular resonances with respect to electric field.
These PRs are significantly different from magnetic Feshbach Resonances (FRs)
in ultracold atomic physics \cite{FR}.
A FR occurs when an external magnetic field separates the scattering
thresholds and alters the molecular structure
and changes the number of bound states in the 2-body system.
This process acts over short range, where spin exchange couples
different channels.  The long range character of the interatomic potential
remains the same in a magnetic field, in contrast to PRs.
  
With the experimental reality of ultracold ground state polar molecules 
rapidly approaching, it is 
necessary to understand both the energy and electric field dependence of 
the scattering. In this paper we obtain total cross sections for the long 
range scattering of RbCs and RbK.  These molecules are considered to 
be in their absolute ground state.  The rest of the paper is structured as 
follows: we briefly review the Stark effect and the dipolar interaction. 
Then results of the scattering are presented with both energy and electric 
field being varied.  We consider the thermally averaged cross section, and 
finally we explore the character of the scattering as a function of energy and 
electric field.

\section{Stark Effect and Molecular Scattering}
We consider the polar molecules to be in their
absolute ground state, including vibration, rotation and
electronic ground state (${}^1\Sigma$).  We assume the molecules
are rigid rotors best described in the $J$ basis, 
$|JM\rangle$ where $J$ is $J=S+L+N$, $S$ and $L$ are the spin and 
orbital angular momentum of the electronic system, and $N$ is the 
rotational state of the molecule.  For these systems $L$ and $S$ are zero. 
$M$ is the projection of $J$ onto the field axis. We ignore the effects 
nuclear spin.

\begin{figure}
\centerline{\epsfxsize=7.0cm\epsfysize=7.0cm\epsfbox{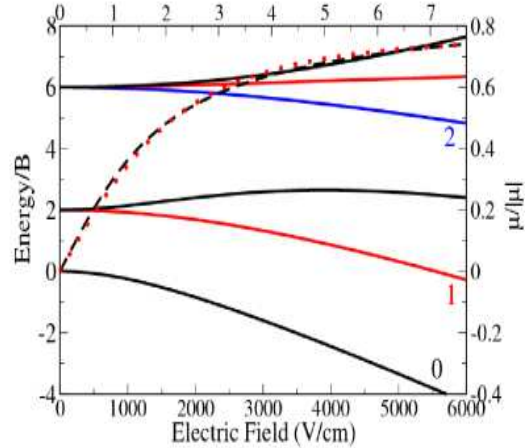}}
\caption{(Color Online) RbCs molecular energies shown as a
function of electric field.  The energies are normalized by the
rotational constant. Different color curves represent 
values of $M$: $M=0$ (Black), $|M|=1$ (Red), and $|M|=2$ (Blue).  
The black dashed line is the projection of the dressed molecular ground 
state onto the field axis, $\langle00|\hat z|00\rangle$, and its value is
given on the right vertical axis.  The red dotted curve is a simple and 
approximate form of $\langle00|\hat z|00\rangle$ given in the text.
The top horizontal axis shows the electric field normalized by the critical 
field value, ${\cal E}/{\cal E}_0$.} \label{starkfig}
\end{figure}

In this model the only molecular structure is the rotational state; with
the electric field accounted for via the Stark effect.
In the $J$ basis the matrix elements of the field-molecule Hamiltonian
and molecular Hamiltonian are written as
\cite{Brown}
\begin{eqnarray}
&&\langle J M |H_{mol} | J^\prime M^\prime\rangle =B
J(J+1)\delta_{JJ^\prime}\delta_{MM^\prime}\nonumber\\ 
&&-\mu_{}{\cal E} [J,J^\prime](-1)^{M}
\left(\begin{array}{ccc}
J&1&J^\prime\\-M&0&M^\prime
\end{array}\right)
\left(\begin{array}{ccc}
J&1&J^\prime\\
0&0&0\end{array}\right),
\label{stark}
\end{eqnarray}
where $[J]$ is a shorthand notation for $\sqrt{2J+1}$.
$B$ is the rotational constant and $\mu$ is the electric dipole moment.
In Fig. \ref{starkfig} the Stark energies for RbCs are shown as a
function of electric field, and the energies are normalized by the
rotational constant. 
At first the energies vary quadratically as the field is
varied, then a transition occurs roughly at ${\cal E}_0 = B/\mu$,
when the Stark energy is roughly equal to the energy rotational
splitting.  Above this field value the energies vary
linearly with respect to electric field.
The top horizontal axis shows the electric field 
normalized by the critical field value, ${\cal E}/{\cal E}_0$.
The different color curves represent different
values of $J$ projected onto the field axis; the values are
$|M|=0$ (Black), $|M|=1$ (Red), and $|M|=2$ (Blue).  The black dashed
line is the projection of the lowest molecular eigenstate of the 
molecule/field Hamiltonian from Eq. (\ref{stark}) or the field-dressed 
ground state onto the
field axis, $\langle00|\hat z|00\rangle=\langle\mu\rangle/|\mu|$.
An approximate polarization is shown in Fig. \ref{starkfig} as the dotted 
red curve.  It is used to derive simple analytic results later.
This approximate form is
\begin{equation}
\langle\mu\rangle\approx
0.78\mu\sqrt{x^2\over6.7+x^2} \label{mu}
\end{equation}
where $x={\cal E}/{\cal E}_0$.  This approximation is within
2\% of $\langle\mu\rangle$ for fields less than 6${\cal E}_0$.

Throughout this paper we use $^{87}$Rb$^{133}$Cs
and $^{87}$Rb$^{41}$K as the example of polar molecules.
For RbCs we use a dipole of $\mu$=1.3 De, a mass of $m$=220 a.m.u. and a
rotational constant of $B$=0.0245 K. For this model the 
critical field value is ${\cal E}_0^{RbCs}\simeq$ 780 $V/cm$.  We also 
consider RbK with the parameters
$\mu$=0.76 De, $m$=128 a.m.u., and $B$=0.055 K \cite{mol}.  This yields a
critical field of ${\cal E}_0^{RbK}\simeq$ 3000 $V/cm$.  

The intrigue of polar molecules is their long range anisotropic 
dipole-dipole interaction 
\begin{eqnarray}
 V_{\mu\mu}&&=-\
{3 (\hat {\bf R} \cdot  \hat{\bf \mu}_1)(\hat {\bf R} \cdot \hat
{\bf \mu}_2)-\hat{\bf \mu}_1 \cdot\hat{\bf\mu}_2 \over R^3},
\nonumber\\ \label{fulldidi}
&&= -{\sqrt{6}\over R^3} \sum_{q} (-1)^q
C^2_{-q} \cdot (\mu_1 \otimes  \mu_2)^2_{q}.
\end{eqnarray}
Here $C^2_{-q}(\theta,\phi)$ is a reduced spherical harmonic that
acts on the relative angular coordinate of the molecules, while 
$(\mu_1 \otimes  \mu_2)^2_{q}$ is the second rank tensor formed from two rank 
one operators which act on the molecular state.  The matrix
elements of the dipole-dipole interaction are:
\begin{eqnarray}
&&\langle J_1M_1J_2M_2 l M_l|V_{\mu\mu}
| J_1^\prime M_1^\prime J_2^\prime M_2^\prime l^\prime M_l^\prime\rangle
\nonumber\\
&&= (-1)^{M_{1}^\prime+M_{2}^\prime+M_{l}+1}
[l,l^\prime,J_1,J_1^\prime,J_2,J_2^\prime]
\nonumber\\
&&\times\left({\mu_{}^2\sqrt{6}\over R^3}\right)
\left(\begin{array}{ccc}l&2&l^\prime\\0&0&0\end{array}\right)
\left(\begin{array}{ccc}l&2&l^\prime\\
-M_l&M_l-M_l^\prime&M_l^\prime\end{array}\right)
\nonumber\\
&&\times\left(\begin{array}{ccc}1&1&2\\
M_{1}-M_{1}^\prime&M_{2}-M_{2}^\prime&M_l-M_l^\prime\end{array}\right)
\nonumber \\
&&\times\left(\begin{array}{ccc}J_1&1&J_1^\prime\\0&0&0\end{array}\right)
\left(\begin{array}{ccc}J_1&1&J_1^\prime\\
-M_{1}&M_1-M_1^\prime&M_1^\prime\end{array}\right)
\nonumber \\
&&\times\left(\begin{array}{ccc}J_2&1&J_2^\prime\\0&0&0\end{array}\right)
\left(\begin{array}{ccc}J_2&1&J_2^\prime\\
-M_{2}&M_{2}-M_{2}^\prime&M_{2}^\prime\end{array}\right).
\label{matdd}
\end{eqnarray}
To perform the scattering calculation we field dress the 
system which entails using the molecular-field eigenstates from 
Eq. (\ref{stark}) to compose molecular states in the scattering channels.  
The basis is also symmetric under interchange because the 
molecules are identical bosons. 

The scattering Hamiltonian can be expressed by the projection of the
total angular momentum, $M_T=M_1+M_2+M_l$, onto the field axis ($\hat z$), 
and $V_{\mu\mu}$ is block diagonal due to rotational 
symmetry about the field axis.  
To obtain the total cross section, $\sigma$, the cross-sections
for each block of total $M$, $\sigma^{(M)}$ must be calculated.
The matrix elements of $V_{\mu\mu}$ differ between
blocks of total $M$ and thus are rigorously required to be computed.
Thus $\sigma$ in terms of $\sigma^{(M)}$ is
\begin{eqnarray}
&&\sigma=\sum_{M}\sigma^{(M)}
\\ \nonumber
&&\sigma_{}^{(M)}=
\sum_{ij}\sigma_{ij}^{(M)}
=\sum_{ij}\frac{2\pi}{k^2}|T_{ij}^{(M)}|^2,\label{sigma}
\end{eqnarray}
where $\sigma_{ij}^{(M)}$ is the cross section for the system to
scattering from the $i^{th}$ to the $j^{th}$ channel for 
$M_T=M$.  $k^2=mE$ where $E$ is the collision energy, and
$T_{ij}^{(M)}$ is the scattering T-matrix with $M_T=M$ \cite{Taylor}.
The factor of $2$ in Eq. (\ref{sigma}) is present because the molecules 
are identical particles. We consider collisions in lowest thresholds with 
two ground state molecules, and there are no 2-body inelastic channels. 
It is also worth noting 
the molecular and scattering Hamiltonian are invariant to the 
sign of the electric field, thus $\sigma^{(-M)}=\sigma^{(M)}$.  

An approximate $\sigma$ can be obtained by assuming $\sigma^{(M)}$ is equal 
to $\sigma^{(0)}$ once all the terms with unphysical partial wave terms have 
been removed, i.e. those with $|M_l|>l$.  To clarify this  
consider $\sigma^{(1)}$ for ground state molecules, $M_1=M_2=0$, and therefore
$M_T=M_l=1$.  Since $M_l=1$ it is unphysical to have s-wave 
channels contribute and these are removed.  This approximation results in  
$\sigma^{(1)}=\sum_{ij}\sigma^{(0)}_{ij}(1-\delta_{l_i0})(1-\delta_{l_j0})$.
Using this procedure for all $\sigma^{(M)}$ we find  
the total approximate cross section is
\begin{eqnarray}
\tilde\sigma=\sum_{ij}(2l_{\min}+1)\sigma_{ij}^{(0)}
\label{app_sigma}
\end{eqnarray}
where $l_{min}=\min(l_i,l_j)$.  This approximation works well and is a cost 
effective method for computing the thermally averaged total cross section.
To obtain the thermally averaged cross section we use
\begin{equation}
\langle\sigma\rangle\approx
\langle\tilde\sigma\rangle={1\over(kT)^2}\int_0^\infty 
E \tilde\sigma(E)e^{-E/kT}dE
\label{thermal}\end{equation}
where $\tilde\sigma(E)$ is the energy dependent approximate total
cross section and $k$ is
Boltzmann's constant and $T$ is the temperature.  

In ultracold atomic collisions there is usually a need to compute only one 
block of $M_T$, the one containing s-wave channels.  
This is due to Wigner threshold laws which state
non-zero partial waves are suppressed as the collision energy goes to zero.
However, this is not the case for polar molecules in a non-zero electric
field.
The electric field mixes the various rotational states and polarizes the 
scattering molecules.  Considering the field dressed scattering
Hamiltonian, we find that there is a direct dipole-dipole coupling between
the scattering molecules.  The result of this interaction is most clearly 
seen through the Born approximation, which assumes a long range potential 
$C_{ij}R^{-s}$.  For small $C_{ij}$ one can approximate 
the wave-functions as spherical Bessel functions and then the Born 
approximation yields a T-matrix \cite{Taylor,threshold}.  
From the Born approximation we obtain a
partial cross section $\sigma_{ij} \propto E^{p}$ where $p=\min(2l,s-3)$.
For a non-zero electric field, we find $s=3$ and $C_{ij}$ denotes the field 
dressed couplings derived from Eq. (\ref{matdd}).  In the Born
approximation when  $C_{ij}\ne0$ the partial cross-section for degenerate 
channels is
\begin{equation}
\sigma_{ij}= constant.\label{born}
\end{equation}
This result is independent of energy for all $l$, and implies the 
dipole-dipole interaction leads to a scattering cross-section where many 
partial waves might contribute even at low energy.  
For a complete discussion of this result see Ref. \cite{threshold}.
 
A worthwhile estimate of the total cross section is with a
semi-classical approach.  This approach offers scaling of $\sigma$
on the physical parameters of the system, such as 
$\mu$, $m$ and $E$ \cite{hydro}.  This yields a total cross section of
\begin{equation}
\sigma_{SC}={\langle\mu\rangle^2\sqrt{m\over E}}c_{SC} \label{hydro}
\end{equation}
where $c_{SC}$ is a constant chosen so the units of 
$\mu$, $m$, $E$, and $\sigma_{SC}$ are $[De]$, $[a.m.u.]$, $[K]$, and 
$[cm^2]$ respectively.  Using Eq. (\ref{mu}) for 
$\langle\mu\rangle$ and comparing the approximation to the 
full scattering calculation with many different initial boundary conditions
at $R_{in}$ with large ${\cal E}$ we obtain $c_{SC}=1.5\times10^{-13}$. 
This is roughly an order of magnitude less that what is obtained from the 
semi-classical calculation \cite{hydro}, but as we shall see this value of 
$c_{SC}$ offers a good representation of the 
scattering in an electric field for both RbCs and RbK.

In quantum mechanical scattering the unitarity limit provides an upper limit
for any single partial wave contribution. This limit is obtained 
when the T-matrix takes on its maximum value of 4, yielding
\begin{equation}
\sigma_Q={8 \pi\over{m E}}c_Q. \label{quantum}
\end{equation}
$c_Q=4.85\times10^{-15}$ and has been chosen so the units of  $m$, $E$, 
and $\sigma_Q$ are $[a.m.u.]$, $[K]$, and $[cm^2]$ respectively.  The 
comparison of $\sigma_{SC}$ and $\sigma_Q$ offers insight into the scattering
process and is explored at length below.  The primary difference between these
two cross sections is the energy dependence, and this indicates there 
will be a transition from semi-classical to quantum mechanical scattering as
the collision energy is lowered.

To numerically solve the scattering problem we use Johnson's log derivative
\cite{Johnson}. We start the scattering calculation at
$R_{in}=20a_0$, which is inside of where the molecular
interaction deviates from $V_{\mu\mu}$ due to 
van der Waals interactions.  At $R_{in}$
we impose the boundary condition that the wave-function must be zero. This 
is not a physical boundary condition; rather it is a starting point to 
systematically study the long range scattering.  We also include a diagonal 
$-C_6/r^6$ potential, where a value of $C_6=10^3$ a.u. is used. We propagate 
the log-derivative to $R_{\infty}=10^5a_0$. To converge $\sigma^{(M)}$ we need a 
large number of partial waves.  For RbCs (RbK) we use $l_{max}=18$ 
($l_{max}=14$).  Furthermore many values of total $M$ are required, 
RbCs (RbK) needs up to $M_T=10$ ($M_T=6$) to converged for
field values up to 4${\cal E}_0$, and collision energies ranging from 
$10^{-7}$ to $10^{-4}K$.
We use up to J=2, with additional rotational states make the calculations very
computationally cumbersome to converge with respect to number of partial wave.
On physical grounds collisions in the ground state are 
coupled at third order to J=3 rotational states and for these reasons are 
omitted.

Using the above parameters we find all $\sigma^{(M)}$s are converged to 
better than 1$\%$ and the total cross section is converged to better 
than 10$\%$ for electric field values up to 4${\cal E}_0$ with
collision energies ranging from $10^{-7}$ to $10^{-4}K$.
For low field and low energy $\sigma$ is converged to a much better percentage.
For energies below $10^{-7}$K we need a larger $R_{\infty}$ to converge 
the calculations, and at higher collision energies, E $>10^{-4}$K, more partial 
waves and $\sigma^{(M)}$ are required.
With this model, we explore the field dependence and energy dependence
of the molecular scattering.
\section{Energy Dependence}
\begin{figure}
\centerline{\epsfxsize=7.0cm\epsfysize=7.0cm\epsfbox{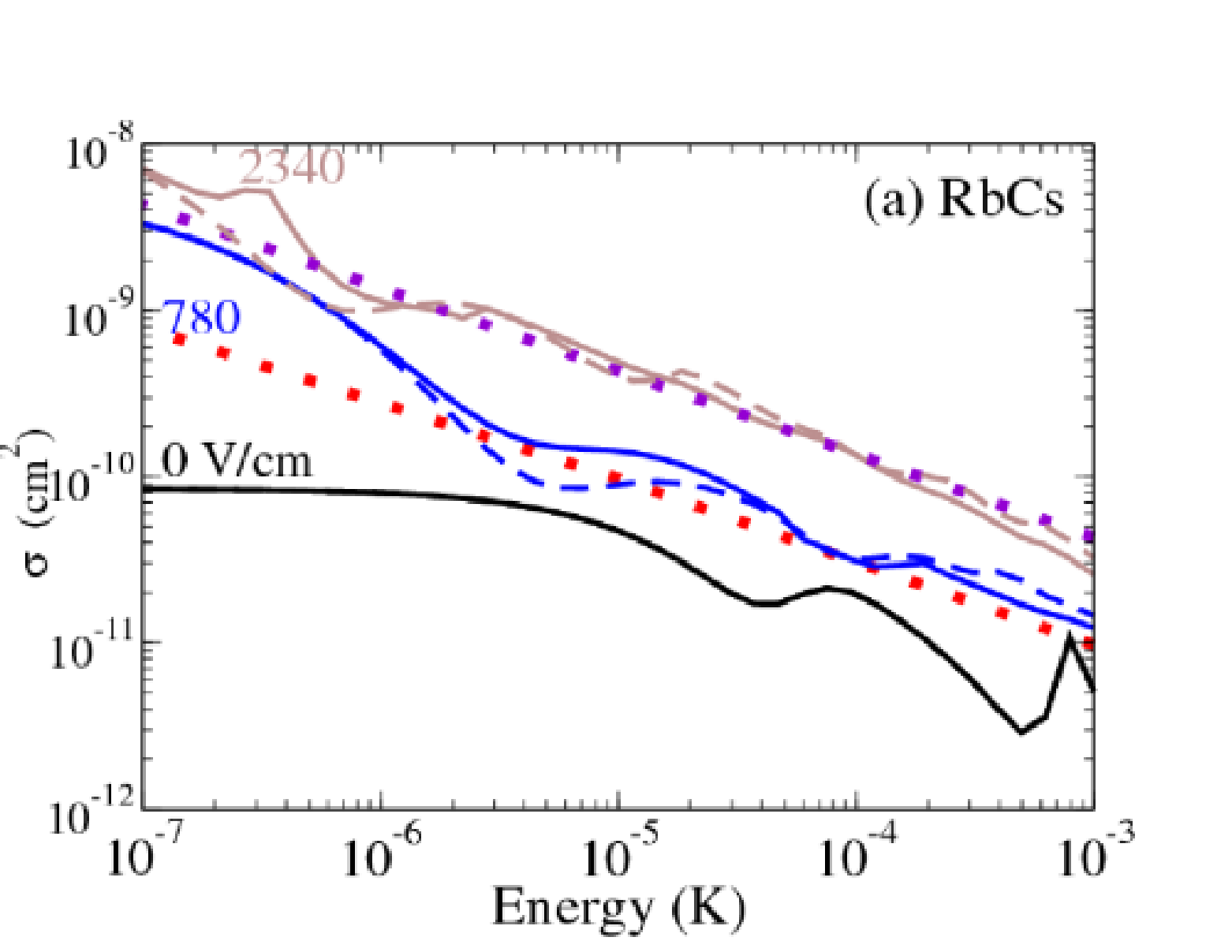}}
\centerline{\epsfxsize=7.0cm\epsfysize=7.0cm\epsfbox{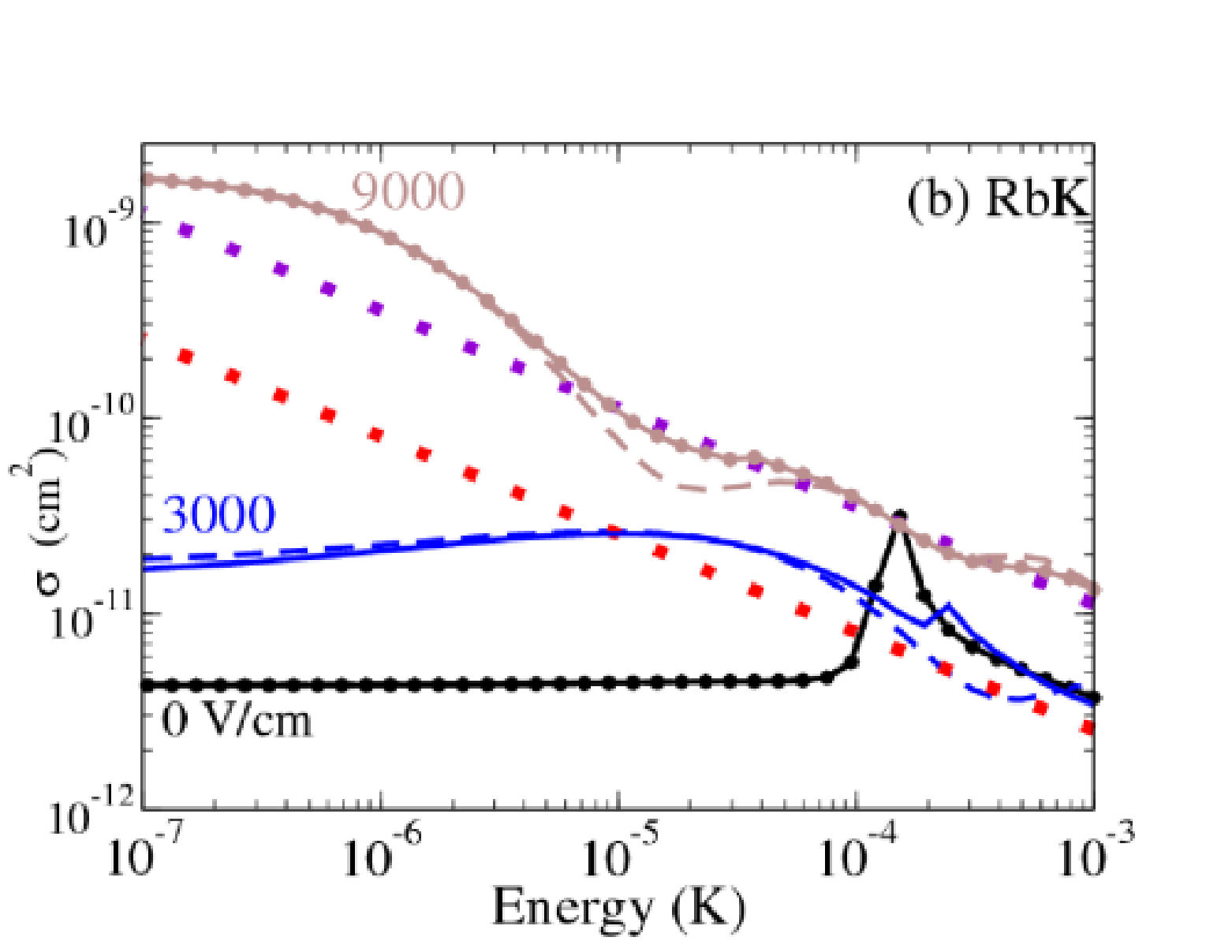}}
\caption{(Color Online) 
The energy dependence of $\sigma$ (solid) and $\tilde\sigma$ (dashed) 
for various electric fields.  In ascending order the fields are 0 (Black), 
${\cal E}_0$ (Blue), and 3${\cal E}_0$(Brown) for (a) RbCs and (b) RbK where
${\cal E}_0^{RbCs}\simeq$ 780 V/cm and ${\cal E}_0^{RbK}\simeq$ 3000 V/cm.
The dotted lines are $\sigma_{SC}$ from Eq. (\ref{hydro}) for different electric
fields in ascending order they are ${\cal E}_0$ (red) 
and $3{\cal E}_0$ (purple).
}\label{energy}
\end{figure}

In Fig. \ref{energy} we show the energy dependence of the total cross section, 
$\sigma$ (solid), and the approximate cross section, $\tilde\sigma$ (dashed),
for both RbCs (a) and RbK (b) at three electric field values: 0 (Black), 
${\cal E}_0$ (blue), and $3{\cal E}_0$ (Brown) 
where ${\cal E}_0^{RbCs}\simeq780$ V/cm and ${\cal E}_0^{RbK}\simeq3000$ V/cm. 
 The dotted lines are $\sigma_{SC}$ from Eq. (\ref{hydro}) for
the electric field values of ${\cal E}_0$ (red) and 
$3{\cal E}_0$(purple).
There are general comments which can be made about both (a) and (b),
but the effects of dipolar scattering are more prominent in the heavier, 
more polar RbCs. 

The solid black curve shows the energy dependent scattering for zero electric 
field.   In zero field there is no dipolar coupling between degenerate 
channels containing ground state molecules.  This fact implies that the
zero field scattering will behave like the familiar ultracold atomic systems,
where the low energy scattering is s-wave dominated.  
This means as the energy goes to zero, $\sigma\rightarrow 8\pi a^2$ where $a$
is the s-wave scattering length.  Furthermore, channels with non-zero partial 
waves are suppressed as the collision energy goes to zero. The zero field 
collisions will contain information about the short range interactions.  
However the dipolar interaction does influence the scattering
at short range when $V_{\mu\mu}$ becomes larger than the threshold separation 
been channels containing two ground state molecules, $J_1=J_2=0$, and 
two rotationally excited molecules, $J_1^\prime=J_2^\prime=1$. 
 
Once there is an electric field, the non-zero partial wave terms 
contribute to the total cross section even at low energy. This can be 
seen in all of the curves in (a) with ${\cal E}\ne0$.  The blue and
brown curves have significantly different profiles than the black curve  
in both (a) and (b).  The change in profile is only slightly due to the 
change in the s-wave scattering length.  Predominantly the change is
due to the additional contribution of non-zero partial waves to the
total cross section.  Generally in a strong field the scattering 
is made up of many partial wave contributions at low energies.  
The total cross section for large electric field is fairly well represented 
by $\sigma_{SC}\propto E^{-1/2}$.  This is seen in (a) and (b) when 
the field is 3${\cal E}_0$, in the similar energy dependence of 
the purple dotted lines and the brown curves.  

At low energy quantum mechanical scattering must dominate, 
$\sigma_Q>\sigma_{SC}$ as $E\rightarrow0$.  This fact does not imply 
that $\sigma$ must be larger than $\sigma_{SC}$ or $\sigma=\sigma_Q$. 
Rather it implies that the scattering will depend on the phase it acquires 
at short range and only when resonant will a single partial wave obtain the 
value of $\sigma_Q$. 

$\tilde\sigma$ from Eq. (\ref{app_sigma}) works well as a cost effective 
method to determine the total scattering cross section.  
This approximation fails when there are resonances in any of the  
$\sigma^{(M)}$s.  If a resonance is in $\sigma^{(0)}$, then $\tilde\sigma$
overestimates $\sigma$ or if there is a resonance in
$\sigma^{(M\ne0)}$ then $\tilde\sigma$ will underestimate $\sigma$.  
Aside from these draw backs $\tilde\sigma$ offers a cost effective method to 
estimate $\tilde\sigma$ over a wide range of energies and electric fields.

\begin{figure}
\centerline{\epsfxsize=7.0cm\epsfysize=7.0cm\epsfbox{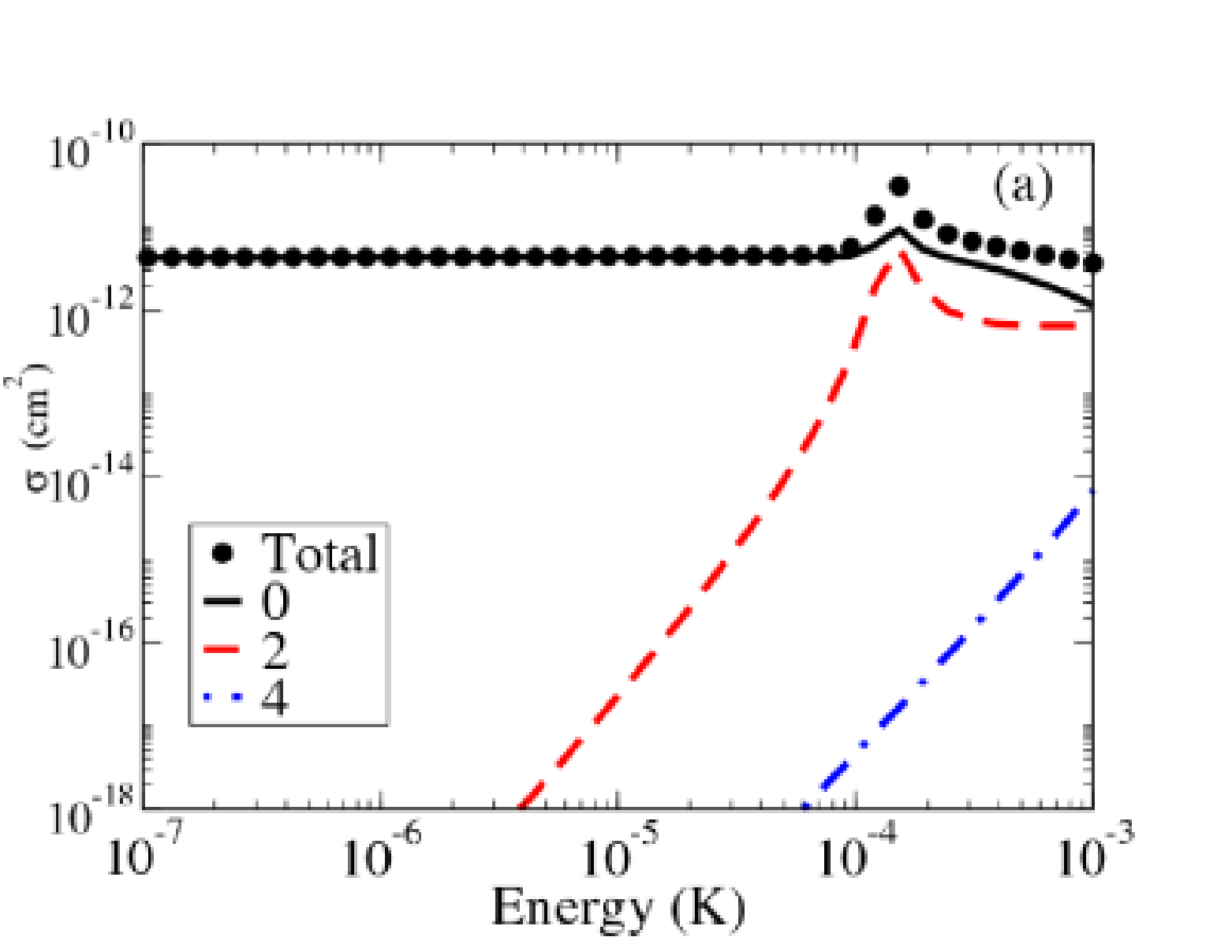}}
\centerline{\epsfxsize=7.0cm\epsfysize=7.0cm\epsfbox{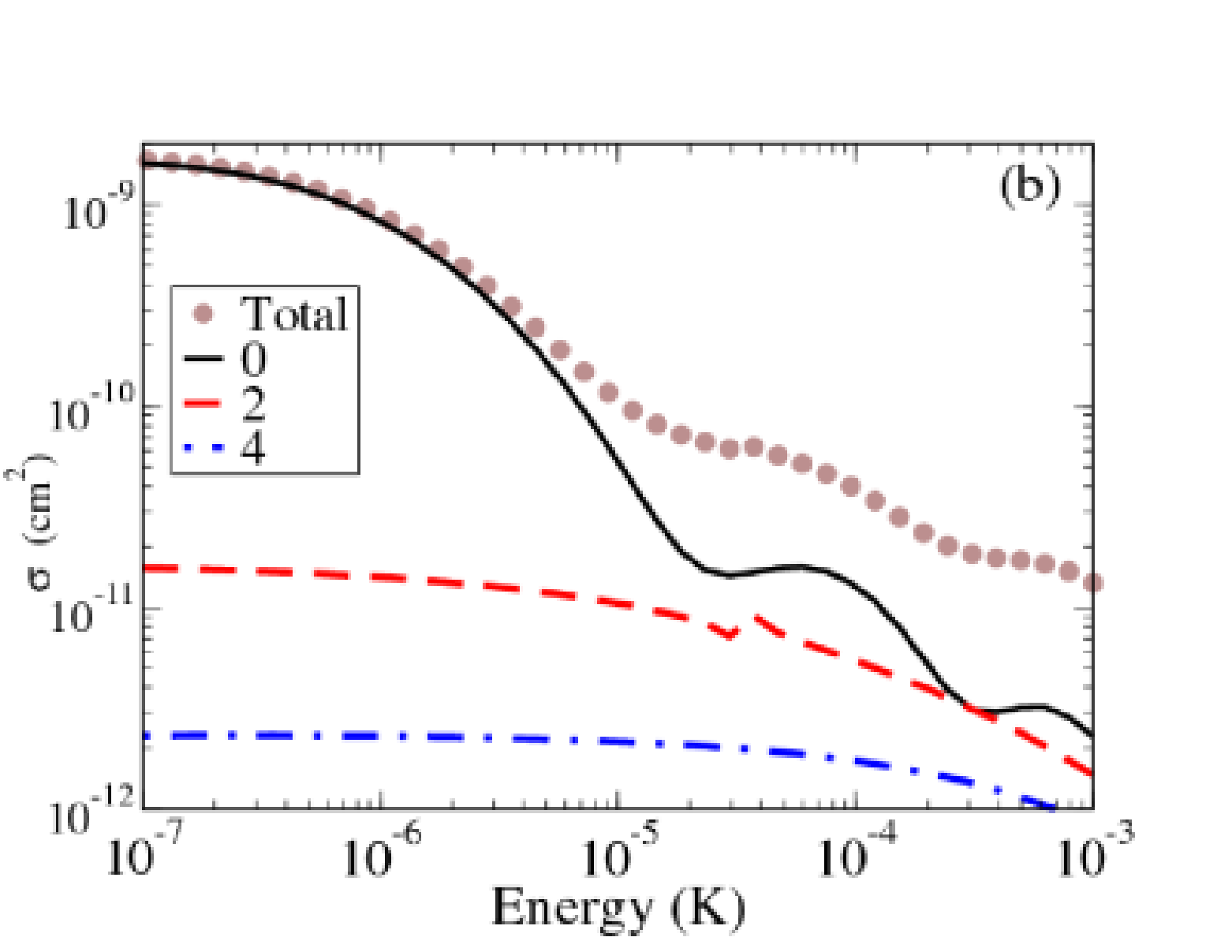}}
\caption{(Color Online) The total cross section (full circles) for RbK 
decomposed into three $\sigma^{(M)}$s for different
field values. The fields are (a) $\cal E$=0 and (b) ${\cal E}=3{\cal E}_0$
and their comparison clearly shows the change in threshold behavior 
induced by the electric field.  The different $\sigma^{(M)}$s are for M 
equal to 0 (solid), 2 (dashed) and 4 (dash-dot).
The total cross section is also shown as circles in Fig. \ref{energy} (a).
}\label{energyterms}
\end{figure}

In Fig. \ref{energyterms} we explore the behavior of particular $\sigma^{(M)}$s
as a function of energy at various electric fields.
(a) and (b) show $\sigma$ and several $\sigma^{(M)}$s 
for RbK at two different field values:  
(a) $\cal E$=0 and (b) ${\cal E}=3{\cal E}_0$.  
The circles are the total 
cross section and are the same points in Fig. \ref{energy} (b).
The  other curves are the different $\sigma^{(M)}$s, where $M_T$ is
0 (solid), 2 (dashed), and 4 (dash-dot).

The total cross section at zero field (black circles) 
and a few of its components are shown in
Fig. \ref{energyterms} (a).  The most important feature
of this figure is that only $M=0$ is significant at low collision energy.
$\sigma^{(0)}$ is the only calculation containing s-wave scattering.
In zero field colliding ground state molecules are  
unpolarized and therefore dipolar interaction is confined to short range.  
This fact results in all non-zero partial waves being suppressed as 
the collision energy goes to zero. This can clearly be seen in the 
$\sigma^{(2)}$ (dashed) and $\sigma^{(4)}$ (dash-dot) where these cross sections 
go to zero as  the collision energy is decreased.  Note the vertical axes 
of (a) and (b) have significantly different scales.

The result of the scattering is significantly different when there is a 
non-zero field.  In Fig. \ref{energyterms} (b) (${\cal E}=3{\cal E}_0$) and 
we have plotted $\sigma$ (brown circles) from Fig. \ref{energy} (b)
and $\sigma^{(0)}$ (solid), $\sigma^{(2)}$ (dashed), 
and $\sigma^{(4)}$ (dash-dot).  In this situation there is direct dipolar 
coupling between two scattering ground state molecules, which means 
these are couplings between degenerate channels in the
field dressed basis.
This coupling dramatically alters the low energy behavior of all non-zero 
partial waves and $\sigma^{(M)}$s; 
they are constant at low energy as predicted by the Born approximation 
in Eq. (\ref{born}).  
These two figures, (a) and (b), show the essential difference between polar 
molecules with and without electric field, and alludes to why 
partial waves and many total $M$s are required to converge $\sigma$, especially
at high collision energies.

We have shown the dramatic effect an electric field has on the
energy dependent scattering.  We now study the scattering of polar
molecules as a function of electric field at various energies and 
temperatures.

\section{Electric Field Dependence}
\begin{figure*} 
\centerline{
\epsfxsize=7.0cm\epsfysize=7.0cm\epsfbox{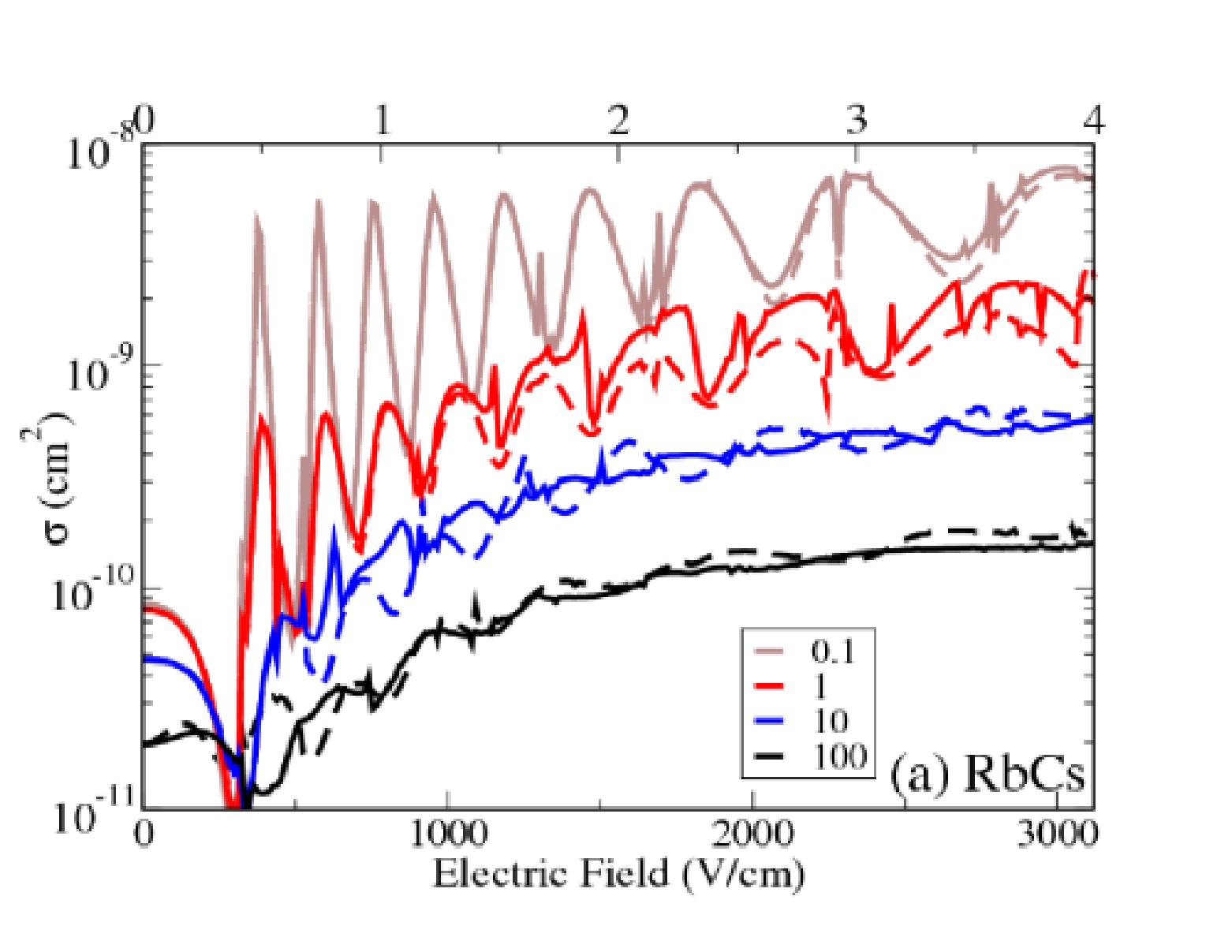}
\epsfxsize=7.0cm\epsfysize=7.0cm\epsfbox{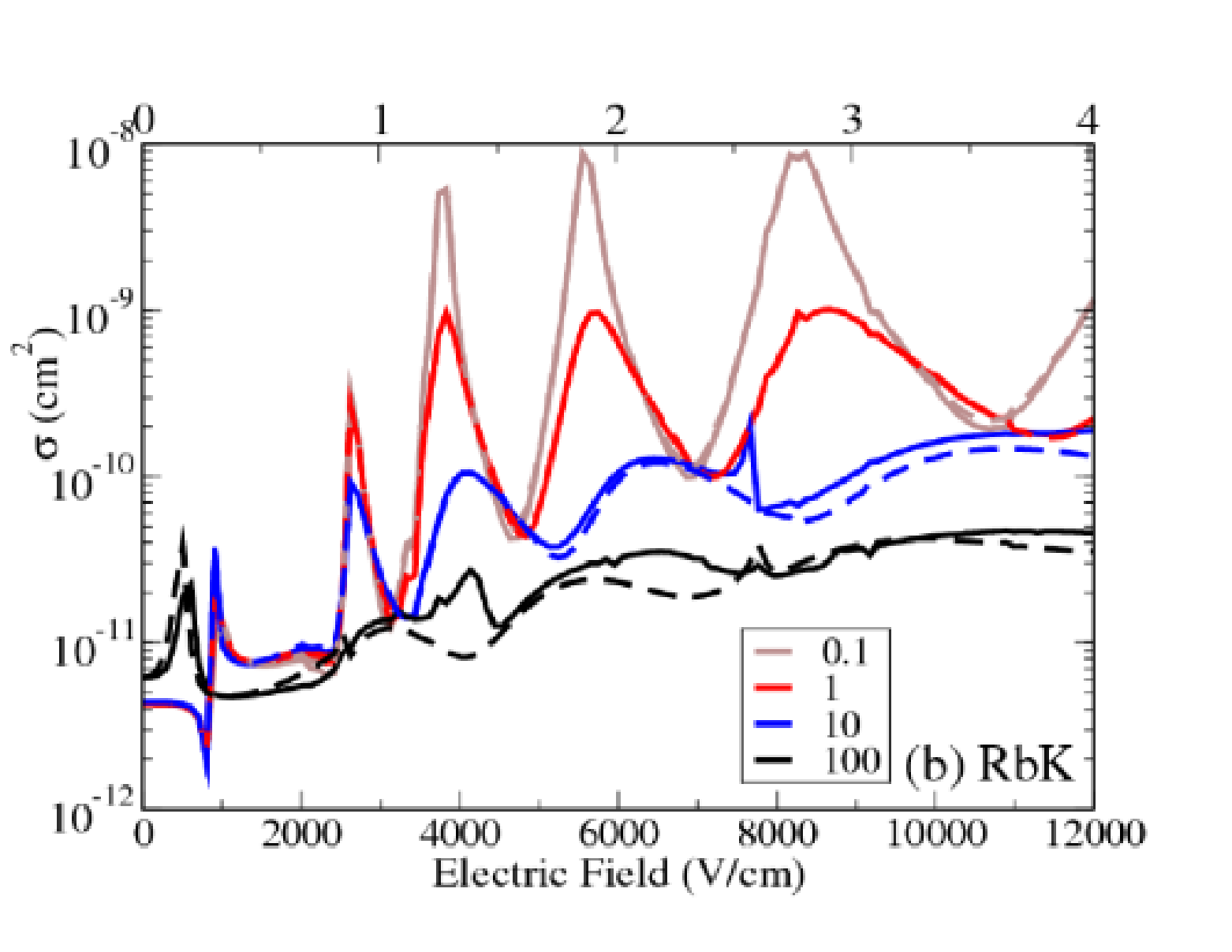}}
\centerline{\epsfxsize=7.0cm\epsfysize=7.0cm\epsfbox{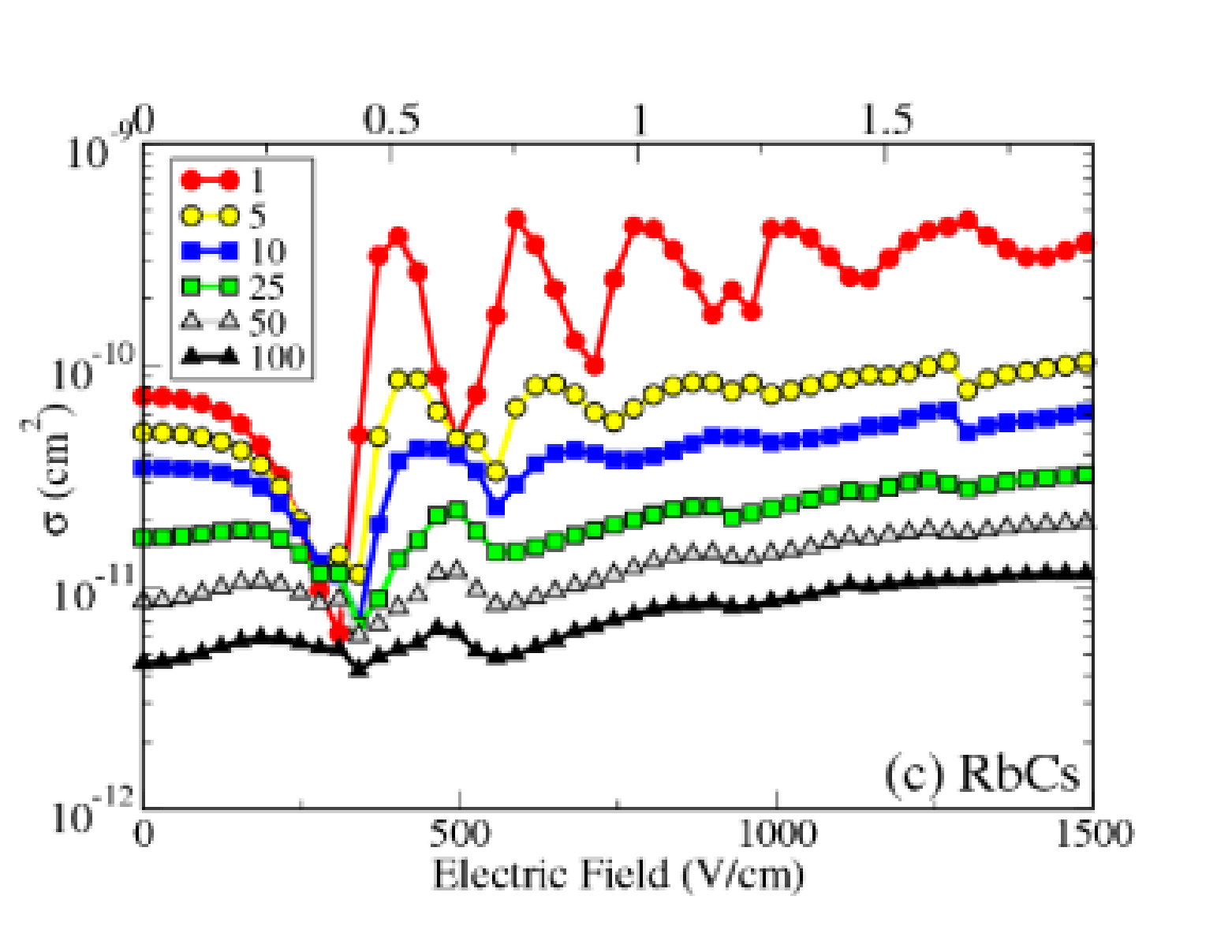}
 \epsfxsize=7.0cm\epsfysize=7.0cm\epsfbox{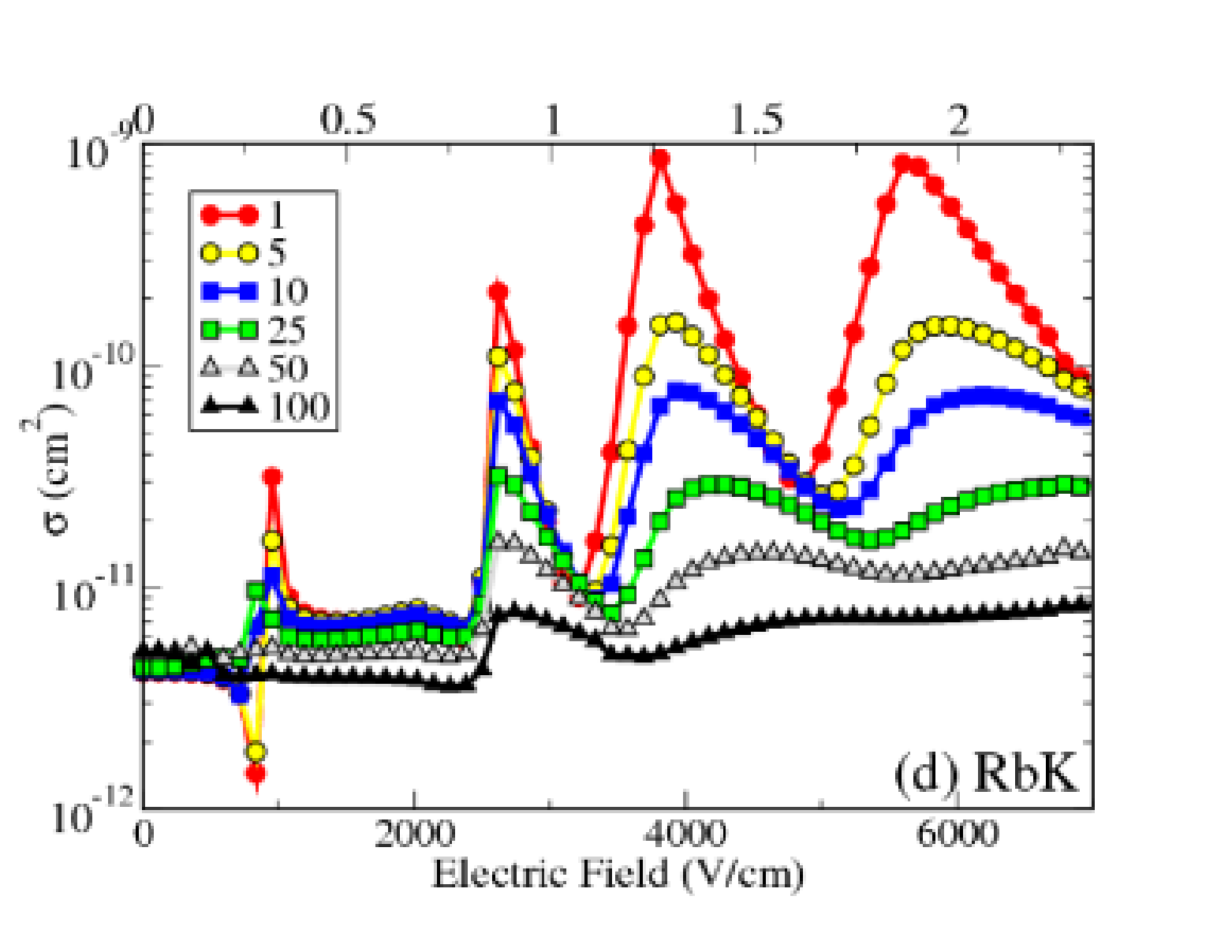}}
\caption{(Color Online) Electric field dependence of the total cross section 
(solid lines) and approximate total cross section (dashed lines) at various 
energies for RbCs in (a) and RbK in (b).  In (a) and (b) the energies of the 
curves in descending order are 0.1 (brown), 1 (red), 10 (blue), and 
100 $\mu$K (Black).  Electric field dependence of the approximate 
thermally averaged cross at various temperatures for RbCs (c) and RbK (d).  
The temperatures of the curves in descending order are 1 (circle), 
5 (open circle), 10 (square), 25 (open square), 50 (open triangle), 
and 100$\mu$K (triangle).}\label{electric}
\end{figure*}

In ultracold atomic physics it is more experimentally
feasible to change an external field rather than the temperature of the gas.  
To this end we study the total cross section and 
approximate total cross section as a function of electric field 
at several different energies and temperatures.  In Fig. \ref{electric} 
(a) and (b) the electric field dependence of $\sigma$ (solid) 
and $\tilde\sigma$ (dashed) are shown for different energy values.
In descending order they are 0.1 (brown), 
1 (Red), 10 (blue), and 100 $\mu$K (black).  
Primarily, the influence of the electric field is to make the cross section 
large and induce potential 
resonances, which are clearly seen in the brown curves in (a) and (b).
The heavier, more polar RbCs has many more than RbK for a field range of
0 to 4${\cal E}_0$.

At the collision energy of 100 $\mu$K (black curves) the dominant effect of 
the electric field is to increase the cross section without pronounced 
PRs.  The result of decreasing the energy an order of 
magnitude (blue) is to make the PRs emerge at low field, but not at high field.
Then at 1 $\mu$K (red) many more PRs become distinguishable and the 
variation in the cross section becomes significant at low field. 
Finally at 0.1 $\mu$K (brown) the PRs are very distinct and there is 
significant variation in $\sigma$ between the PRs at low field. 
At high field, especially in RbCs, the minima between PRs are less deep.
The decrease in variation of $\sigma$ is clearly seen in the brown curve 
in both (a) and (b).  This result is simply due to the larger number of 
partial waves contributing to the scattering cross section.  Ultimately,
the electric field does not offer control of the scattering length as has
been seen with the magnetic Feshbach resonance.  Rather its induces a large 
number of partial waves to significantly contribute to $\sigma$, thus resulting
in large total cross sections, but they are not necessarily resonant.

In a magnetic Feshbach resonance the field alters the molecular structure 
so the colliding pair of atoms can access an alternate pathway (closed
channel quasi-bound state). The pathways can interfere constructively or 
destructively depending on the value of the magnetic field, and this leads
to the ability to tune the scattering.
 
The possibility of resonantly ``turning off'' the 2 body interactions 
in a system of polar molecules with an 
electric field does not truly exist.  The electric field might effectively
turn off the 2 body interaction if the zero field scattering length is greater 
than zero.  This can be seen 
in  Fig. \ref{electric} (a).  This system  and its particular parameters result
in a large positive scattering length ($\sim350$ $a_0$).  This results in a 
minimum in the cross section as the electric field evolves the system 
toward the addition of another bound state.  This fact offers a simple
means to determine the sign of the zero field scattering length by varying the 
electric field.  
Overall, the effect of an electric field is to activate the dipoles and 
make many partial waves significant in the scattering.  This fact prevents 
the cross section from rigorously being zero due to the contribution of 
non-zero partial wave.

In both  \ref{electric} (a) and (b) we see that $\tilde\sigma$ (dashed) 
offers a good approximation to $\sigma$ (solid).  It works especially well 
at low field and low 
energy.  Even at high field and high energy it offers a reasonable estimate 
of the total scattering cross section.  Also seen in Fig. \ref{energy}, 
$\tilde\sigma$ does not get the resonant values of $\sigma$, but still 
offers a cost effective means to achieve a total cross section.  We have
used $\tilde\sigma$ to estimate a thermally averaged total cross section 
at low field, and this is shown in Fig. \ref{electric} (c) for RbCs and (d) 
for RbK.  The temperatures of the curves  are 
1 (circle), 5 (open circle), 10 (square),
25 (open square), 50 (open triangle), and 100$\mu$K (triangle).

The thermally averaged cross section for both RbCs (c) and RbK (d) show that 
as the temperature is lowered from 100 $\mu$K resonant features emerge in 
the cross section. When the temperature is decreased, the RbCs cross section 
develops PRs at low field first. One can clearly see the emergence of the
PRs as the temperature is decreased at low field along with the minimum.
This shows that if the zero field scattering length is greater than zero, 
there will be an observable minimum before the first PR below 
$T\sim$ 25 $\mu$K.  Then as the temperature is further decreased the other 
PRs at high field become distinguishable.
In the thermally averaged system the dominant effect of increasing
the electric field is to raise the total values of the cross section.

\section{Transition in Character of scattering}
\begin{figure}
\centerline{\epsfxsize=6.50cm\epsfysize=6.cm\epsfbox{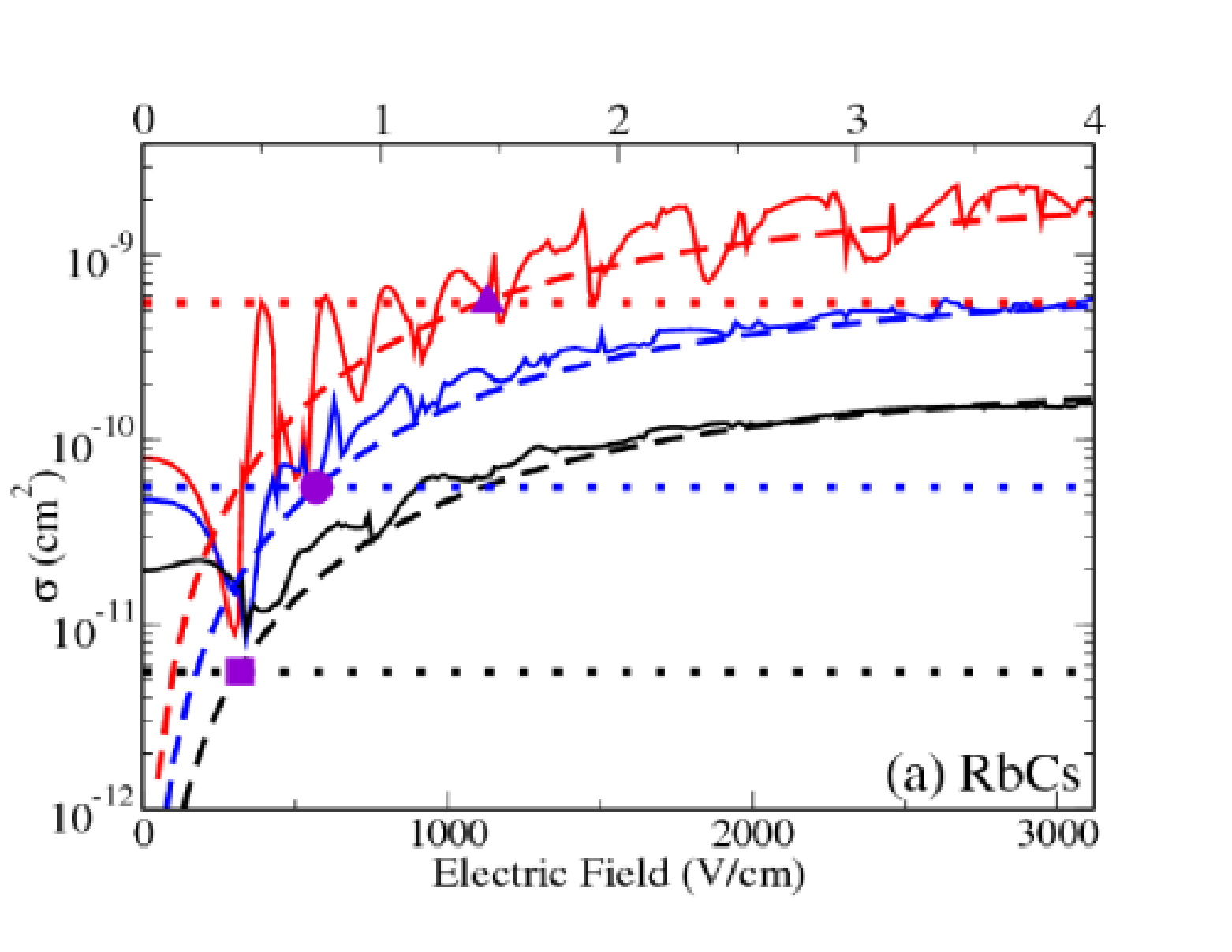}}
\centerline{\epsfxsize=6.50cm\epsfysize=6.cm\epsfbox{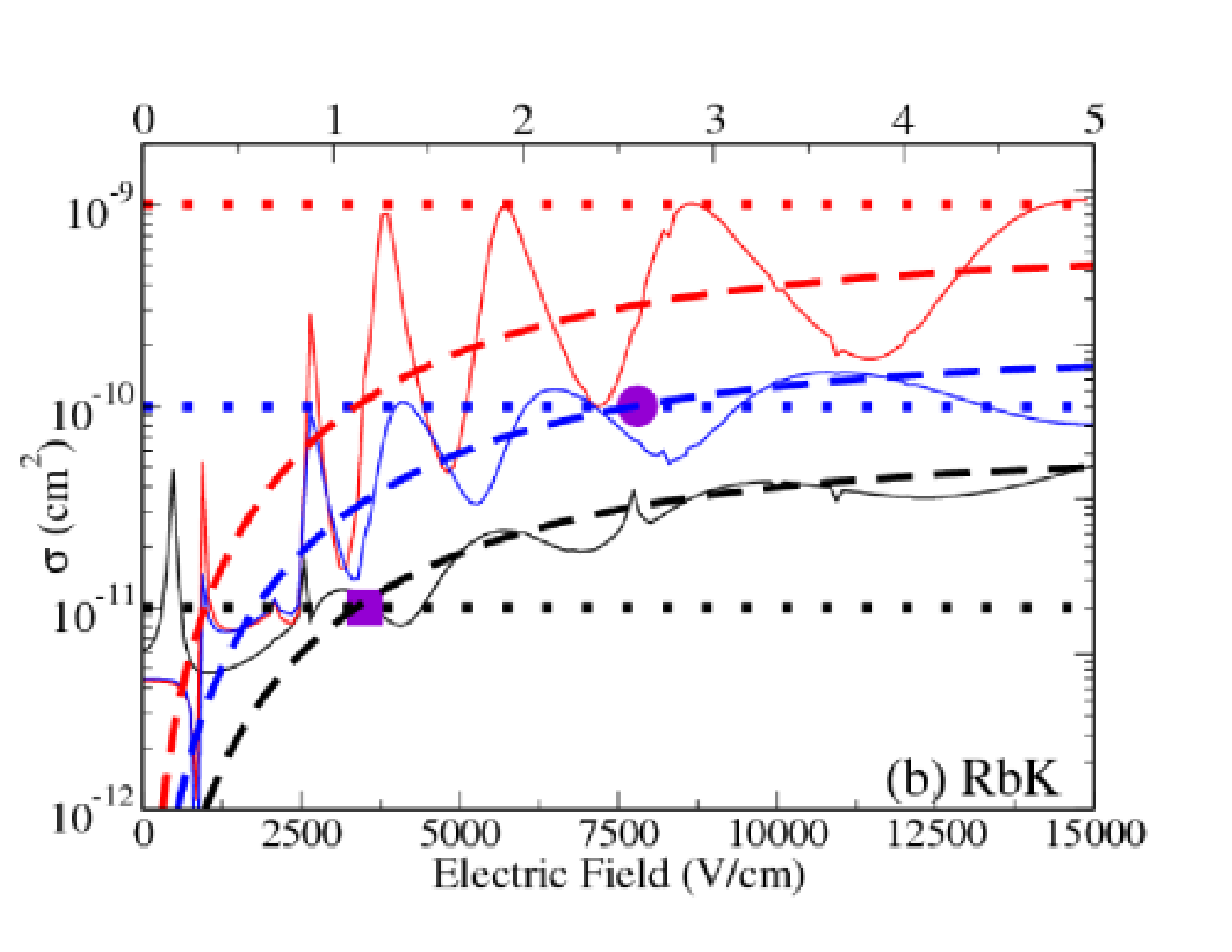}}
\centerline{\epsfxsize=6.50cm\epsfysize=6.cm\epsfbox{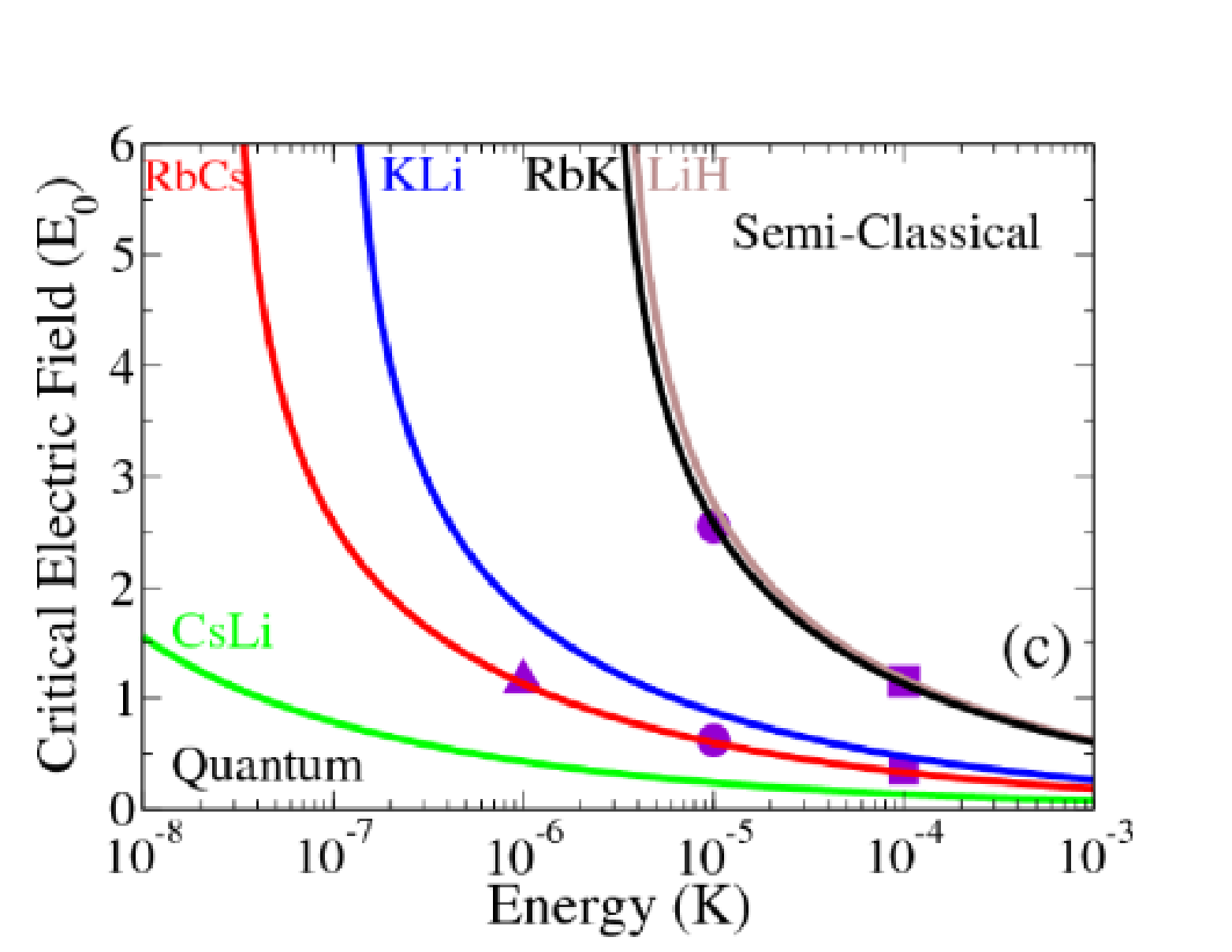}}
\caption{(Color Online) 
Comparisons of $\sigma$ (solid), $\sigma_{SC}$ (dashed) and $\sigma_Q$ (dotted)
show the change of character in scattering for both RbCs (a) and RbK (b).  
Sets of curves ($\sigma$, $\sigma_{SC}$, and $\sigma_{Q}$) are given for 
different energies and in descending order the sets are for 1 (red), 10 (blue) 
and 100 $\mu$K (black).  The critical electric field, when 
$\sigma_{SC}=\sigma_Q$, is indicated by a symbol for each energy, and these
symbols are also shown in (c).  
(c) The critical electric field in units of ${\cal E}_0$ is plotted as a 
function of energy for various molecules. For a particular molecule, 
the scattering is semi-classical in character when the electric field is
greater than ${\cal E}_X$ and the collision energy is greater than $E_Q$.
The scattering is quantum mechanical when either the electric field is less
than ${\cal E}_X$ or the collision energy is less than $E_Q$.
The parameters of the molecules are given in the text.}\label{trans}
\end{figure}
To understand the scattering of polar molecules, we compare the total cross 
section, semi-classical cross section, and quantum unitarity limit scattering 
cross sections.  We make this comparison for both RbCs in Fig. \ref{trans} (a) 
and for RbK in (b).  In each plot we show all three cross sections,
$\sigma$ (solid), $\sigma_{SC}$ (dashed), and $\sigma_Q$ (dotted), at three
energies: 1 (red), 10 (blue) and 100 $\mu$K (black).
To make $\sigma_{SC}$ simple to use Eq. (\ref{mu}).

There is an intriguing interplay between the energy and electric field in this
system.  Equations (\ref{hydro}) and (\ref{quantum}) show us  
$\sigma_{SC}\propto\frac{\langle\mu\rangle^2}{\sqrt{E}}$ and
$\sigma_Q\propto\frac{1}{E}$.  
When $\sigma_{SC}$ is larger than $\sigma_{Q}$, the $\sigma$ will
be made up of a large number of partial waves and have roughly the same simple
energy and electric field dependence of $\sigma_{SC}$. In contrast
to when $\sigma_{Q}$ is larger than $\sigma_{SC}$,
the scattering will sensitively depend on the scattering process which occurs, 
making the energy and electric field dependence non-trivial.  

To begin the analysis we look at RbCs in Fig. \ref{trans} (a). 
First we look at the 100 $\mu$K system: $\sigma$ is the solid black curve 
and its behavior closely follows $\sigma_{SC}$, the dashed black line.
Note how large the cross section becomes for large electric field values.
It is much larger than $\sigma_Q$, the dotted black line.  This shows 
a large number of partial waves are contributing to the scattering.
The electric field where $\sigma_{SC}=\sigma_Q$ is marked by a square
and is called the critical field, ${\cal E}_X$. Above this field $\sigma$
closely follows $\sigma_{SC}$.

When the energy is lowered by an order of magnitude to 10 $\mu$K, 
$\sigma_Q$ (blue dotted) is larger by an order of magnitude, but 
$\sigma_{SC}$ (blue dashed) only increases by a factor of
$\sqrt{10}$. For this system, the  $\sigma_{SC}$ quickly becomes 
larger than $\sigma_Q$ as the electric field is increased and 
${\cal E}_X$ is marked by a circle for this energy.  Above ${\cal E}_X$
we again see $\sigma$ closely follows $\sigma_{SC}$.

At 1 $\mu$K we see there are many potential resonances before 
${\cal E}_X$, marked by a triangle.
Below the critical field $\sigma$ (red solid) has a series of potential 
resonances which reach up to $\sigma_Q$ (red dotted).
Then as the electric field is increased near and above ${\cal E}_X$,
$\sigma$ becomes larger than $\sigma_Q$ and roughly follows the trend of
$\sigma_{SC}$ (red dashed).  It is worth noting that $\sigma$ has many 
fluctuations due to potential resonances in many partial waves 
and $\sigma^{(M)}$s.  

Now turning our attention to RbK, we have plotted 
$\sigma$ (solid), $\sigma_{SC}$ (dashed), and $\sigma_Q$ (dotted) 
for the same collision energies, 1 (red), 10 (blue), and 100 (black) $\mu$K.  
There are some similarities to 
RbCs, but there is a very important difference in the behavior of the 
scattering cross section at 1  $\mu$K: for all electric fields 
$\sigma_Q>\sigma_{SC}$.  
This fact brings up an important quantity $E_Q$, the energy at which 
$\sigma_Q=\sigma_{SC}({\cal E}\rightarrow\infty)$ or for a maximally 
polarized molecule $\sigma_{SC}$, $E_Q$ 
is the energy at which $\sigma_Q$ equals $\sigma_{SC}$.  For collision 
energies below $E_Q$ the scattering will always be dependent on the details 
of the interaction, $\sigma_Q>\sigma_{SC}$.

To explore Fig. \ref{trans} (b) more thoroughly we look at the cross sections
for 100 $\mu$K.  We see that $\sigma$ (solid black) roughly follows
$\sigma_{SC}$ (dashed black) when the field is greater than ${\cal E}_X$.  
Here many partial waves are contributing to the 
scattering.  Again the critical field is marked by a circle and square
for both 10 and 100 $\mu$K, respectively.
As the collision energy is decreased the variation in the cross section 
as a function of electric field is greater.  This can be seen in both 
1 and 10 $\mu$K cross sections. At 1 $\mu$K for all electric fields 
$\sigma_Q$ (dotted red) is greater than $\sigma_{SC}$ (dashed red), 
because $E<E_Q$.
Satisfying the inequality $\sigma_Q>\sigma_{SC}$ does not imply that 
$\sigma>\sigma_{SC}$ or  $\sigma=\sigma_{Q}$.  Rather it signifies the 
scattering will be sensitive to the scattering processes which 
occur.  Thus when the scattering is in the quantum mechanical regime it 
depends on the short range details of the system and will exhibit 
resonance behavior.  This is in contrast to when $\sigma_{SC}>\sigma_{Q}$. 
When this inequality is true, we expect  $\sigma\sim\sigma_{SC}$
as is shown in \ref{trans} (a) and (b).

To explore the interplay of $\sigma_{SC}$ and $\sigma_{Q}$ we analytically 
determine ${\cal E}_X$ as a function of energy.  The critical field is of 
fundamental importance because it denotes the field near which the character 
of the scattering changes from interaction sensitive to semi-classical.  
Equating  $\sigma_{SC}({\cal E})$ and $\sigma_{Q}$ and solving for
the critical field we find:  
\begin{equation}
{{\cal E}_{X}\over {\cal E}_0}=\sqrt{6.7 b\over \sqrt{m^3\mu^4E}-b}
\label{xfield}
\end{equation}
where is $b=\left(8\pi c_Q/0.608c_{SC}\right)$=1.29 and $m$, $\mu$, and 
$E$ are in units of [a.m.u.], [De], and [K], respectively.
We have plotted this critical field in Fig. \ref{trans} (c) for many different 
molecules including LiH (brown), RbK (black), KLi (blue), RbCs (red), and 
CsLi (green). The parameters used for this figure are listed below.
\newline\newline
\centerline{
\begin{tabular}{ccccc}
Molecule& $\mu$ [De]&${\cal E}_0$[V/cm] & $m$ [a.m.u.]&$E_Q$[$\mu K$]\\
\hline
LiH &5.88 &77500&8&2.7\\ 
RbK&0.76&3000&128&2.4\\
KLi& 3.53 &4320&48&0.1\\
RbCs&1.30&780&220&0.02\\
CsLi&5.51& 1850& 140&0.0007\\ \hline
\end{tabular}}
\newline\newline
This figure has interesting features; the most important of which is that it
divides the electric field-energy parameter space into 2 regions which have
qualitatively different scattering character.  Above the curve with a large 
electric field or high energy, the scattering is semi-classical and
the scattering is essentially determined by physical parameters of the 
system: $m$, $E$, and $\langle\mu\rangle$ as shown in Eq. (\ref{hydro}).
Below the curve, for low energy or low electric field, the molecular
scattering is sensitive to the details of the interaction and 
will be characterized by resonances and large variations 
in $\sigma$.  
 
The two regions are labeled such that above the curves the scattering is 
semi-classical and below the scattering is quantum mechanical.
Relating these curves back to the total cross section shown in 
Fig. \ref{trans} (a) and (b), we have included the symbols
in (c).  This offers a feel for how the character changes
above and below ${\cal E}_X$ for a few energies.

The energy below which all scattering is quantum mechanical, $E_Q$, is 
determined when 
$\sigma_Q=\sigma_{SC}({\cal E}_X\rightarrow\infty)$.  This energy is most 
easily determined as the denominator on the right hand side of 
Eq. (\ref{xfield}) goes to zero and is
\begin{equation}
E_{Q}={b^2\over m^3\mu^4}. 
\end{equation}
This reveals for heavier and more polar molecules that the semi-classical 
scattering will occur at a lower energy, suppressing quantum mechanical 
scattering.  This is most evident in Fig. \ref{trans} (c) for CsLi, a 
heavy and very polar molecule, whose $E_{Q}$ is 0.7 $n$K compared with the
polar but light LiH for which $E_{Q}$ is 2.7 $\mu$K or the not very polar 
RbK for which $E_{Q}$ is 2.4 $\mu$K.  It 
is important to notice that the scaling of  $E_{Q}$ is $m^{-3}$ and 
$\mu^{-4}$.  This shows that the mass of the molecule plays a significant role,
almost as significant as the dipole, in determining the character of the 
scattering as the collision energy it lowered. 

To show polar molecules present a unique opportunity to study semi-classical 
scattering consider Chromium 52 \cite{dicold}.  This system has a magnetic 
dipole moment of 6 Bohr magneton.  If we make the appropriate conversions 
for this magnetic dipole moment and put it into the current theory, we find 
$E_Q\sim1$K. This shows the scattering will always be quantum mechanical 
($E<<E_Q$) and there will never be scattering of a semi-classical character 
in this atomic system.  

\section{Conclusion}
We have studied the collisions of RbCs and RbK in an electric field 
at various energies and temperatures.  
This work found the full scattering cross section is a computationally intense 
calculation where a large number of partial waves 
and many blocks of total $M$ are required to converge
the total cross section.  An approximate cross section is introduced and 
works well to provide a cost effective method to obtain a thermally averaged 
cross section.  At large electric fields and at relatively high energies the 
semi-classical scattering cross section approximates the total cross section
well.  

This work has illustrated how dipolar interactions alter the scattering.  
The most notable is that an electric field non-zero partial wave
cross section does not go to zero as the scattering energy goes to zero.
Furthermore resonant control of the interaction will only exist to 
a limited extent for polar molecules.  The electric field in general
induces large total cross sections.  It cannot be used to ``turn off'' the
2 body interactions unless the zero field scattering length is greater than 
zero.

The primary finding of this work is that scattering can be classified as
semi-classical and quantum mechanical.  Semi-classical scattering is 
relatively simple where the scattering is determined by the scattering 
energy ($E$), the molecular mass ($m$), and {\it induced} dipole moment of 
the molecule ($\langle\mu\rangle$) as shown in Eq. (\ref{hydro}).  
Quantum mechanical scattering is behavior defined by resonantly large 
cross sections and is sensitive to the phase shift acquired by the scattering 
process.  We have found a simple form of the critical electric field 
(${\cal E}_X$)  at which the character of the scattering changes as a function 
of energy.  Additionally we have found the collision energy ($E_Q$) below which
all scattering will be quantum mechanical.

An exciting feature of Fig. \ref{trans} is one can study 
the transition of a gas from quantum mechanical to semi-classical scattering, 
simply by turning on an electric field.  This transition in scattering 
character might signify a phase transition as was recently suggested for a 
2 dimensional gas of polar molecules when their interactions have been
modified with microwave fields \cite{buchler}.  

\begin{acknowledgments}
The author is grateful for support from the ARC through ACQAO and computing 
resources from Victorian Partnership for Advanced Computing (VPAC) and the
Centre for Astrophysics and Supercomputing at Swinburne University.
The author thanks Peter Hannaford and Andy Martin for critically reading 
the manuscript.
\end{acknowledgments}
$^\dagger${cticknor@swin.edu.au}
\bibliographystyle{amsplain}

\end{document}